\documentclass[conference]{IEEEtran}
\IEEEoverridecommandlockouts
\usepackage{cite}
\usepackage{amsmath,amssymb,amsfonts}
\usepackage{algorithmic}
\usepackage{graphicx}
\usepackage{textcomp}
\usepackage{xcolor}
\usepackage{soul}
\def\BibTeX{{\rm B\kern-.05em{\sc i\kern-.025em b}\kern-.08em
    T\kern-.1667em\lower.7ex\hbox{E}\kern-.125emX}}
\usepackage{fancyhdr}
\fancypagestyle{eCF}{%
\fancyhf{}
\fancyfoot[L]{eCF Paper Id: 574079}

}
\begin{document}

\title{Collective electrical response of simulated memristive arrays using SPICE\\ 
}

\author{\IEEEauthorblockN{G. A. Sanca, F. Di Francesco}
\IEEEauthorblockA{\textit{Escuela de Ciencia y Tecnolog\'ia} \\
\textit{Universidad de San Mart\'in}\\
San Martín, Argentina \\
gsanca@unsam.edu.ar      fdifrancesco@unsam.edu.ar}
\and
\and
\IEEEauthorblockN{F. Golmar, C.P. Quinteros}
\IEEEauthorblockA{\textit{Escuela de Ciencia y Tecnolog\'ia} \\
\textit{Universidad de San Martín, CONICET}\\
San Martín, Argentina \\
fgolmar@unsam.edu.ar         cquinteros@unsam.edu.ar}
}

\maketitle

\begin{abstract}
Self-assembled structures are possible solutions to the problem of increasing the density and connectivity of memristive units in massive arrays. Although they would allow surpassing the limit imposed by the lithographic feature size, the spontaneous formation of highly interconnected networks poses a new challenge: how to characterize and control the obtained assemblies. In view of a flourishing field of such experimental realizations, this study explores the collective electrical response of simulated memristive units when assembled in geometrically organized and progressively distorted configurations. 
We show that highly idealized memristive arrays already display a degree of complexity that needs to be taken into account when characterizing self-assemblies to be technologically exploited. Moreover, the introduction of simple distortions has a considerable impact on the available resistance states and their evolution upon cycling. Considering arrays of a limited size, we also demonstrate that the collective response resembles aspects of the individual model while also revealing its own phenomenology.
%
\end{abstract}


\vspace{0.2cm}

\thispagestyle{eCF}

\noindent In recent years, arrays of memristive devices are attracting more attention due to their versatility to introduce a \mbox{variety} of novel and promising technological implementations~\cite{chua_handbook_2019}. \mbox{Originally} claimed to be an alternative to non-volatile information storage\cite{li_memristive_nodate,zahoor_resistive_2020},
they are now at the focus of an intense research effort to develop neuromorphic circuits~\cite{hochstetter_avalanches_2021}. Since memristive properties have been assimilated to particular synapses' and neurons' electrical responses~\cite{sah_brains_2019,chua_memristor_2019,cai_synapse_2019}, assembling memristive units into a statistically representative number appears as an inevitable strategy to compare them to their biological counterparts~\cite{chialvo_emergent_2010}. Moreover, it has been proposed that a large enough collection of synaptic weights could demonstrate emergent behavior capable of partially projecting in hardware operations that, otherwise performed in software, are complex and inefficient~\cite{wang_deep_2019,karbachevsky_early-stage_2021}. The key to unveiling these phenomena is to increase the number of units in the system.   

Experimentally, achieving the referred statistically representative number of components~\cite{chialvo_emergent_2010} - and consequently the expected degree of connectivity among them - requires either the most advanced lithography tool (in order to pack as many units as possible) or alternative strategies, such as promoting the coalescence of self-assemblies (that are spontaneously formed collections of multiple units). The latter includes systems such as nanowire networks~\cite{diaz-alvarez_emergent_2019}, core-shell assemblies~\cite{fabris_tunnel_2019}, or domain wall patterns in ferroic systems~\cite{catalan_domain_2012}. Nevertheless, beyond the undeniable convenience of exploiting such a self-formed collection of interconnected units, their electrical characterization has to be reformulated since access to each specific unit is no longer available.

The purpose of this work is to demonstrate, by means of a simulation platform, that simple assemblies of identical and reliable memristive units already display a plethora of collective electrical responses. This, in turn, implies there exists complexity as a feature of those assemblies, implemented as distorted geometrical arrays. Even though complexity is a desired characteristic for these artificial implementations, since it resembles their biological counterparts, it also means that the electrical response of those array gets conditioned by the history of applied stimuli. The term `collective' is key to understand the precise approach considered within this study. A global parameter named after collective resistance is used to identify the changes suffered by arrays of memristive units quantifying the electrical response using only two accessing ports. This approach is motivated by the implementations of the aforementioned self-assemblies. In that case, isolating the individual constituents is not possible and electrically characterizing them necessarily implies averaging the response across multiple units as an inevitable side effect of the high degree of connectivity intrinsically achieved. Following a previous communication~\cite{fdf2021}, in which the spatiotemporal evolution of a homogeneous array is demonstrated, the present work aims at carefully studying the effect on the arrays' collective resistance of different external stimuli, describing the evolution of the global response upon insistence, and highlighting changes when the originally homogeneous array is distorted.

\begin{figure*}[ht!]
    \centering
    \includegraphics[width=18 cm]{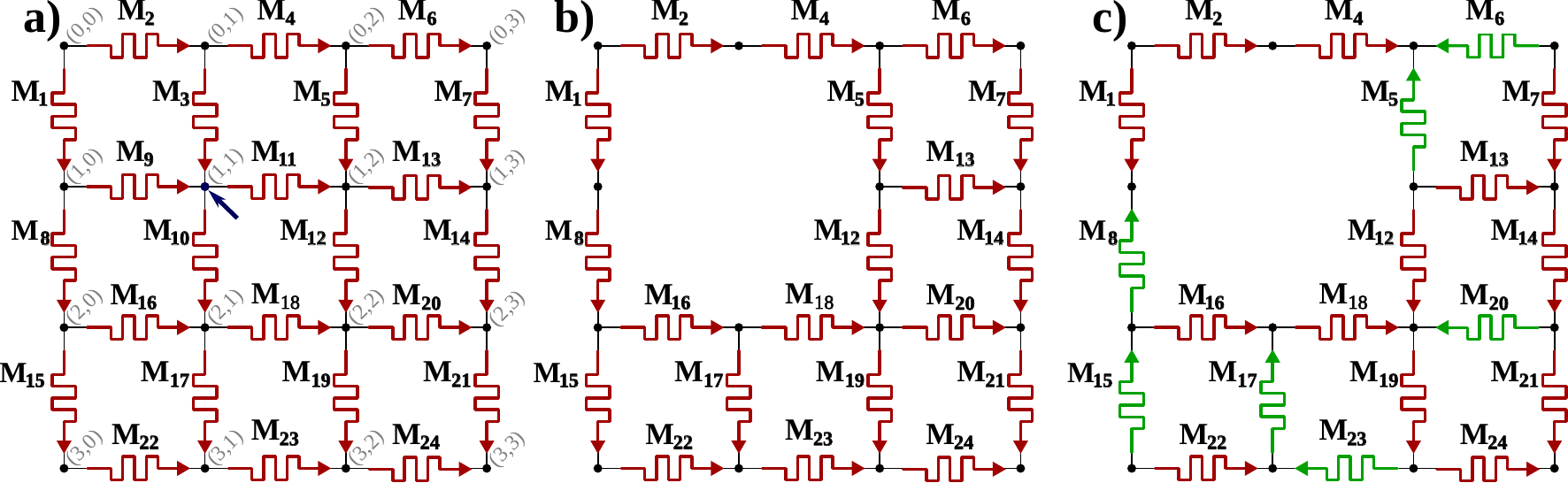}
    \caption{Algorithm progression from the so-called homogeneous array to a distorted one. \textbf{a)} Schematic diagram of the homogeneous case, including nodes' and memristive units' designations. Since the memristive model is bipolar, the red arrows specify \textit{by-default} polarity. 
    \textbf{b)} Distorted array upon node removal (p$_r \neq 0$). After setting the original homogeneous array, the algorithm computes node by node removal probability based on the p$_r$ factor. In \textbf{b)}, the node indicated by the blue arrow in \textbf{a)} is removed, eliminating all the devices connected to it.  
    \textbf{c)} Further distorted array allowing polarity inversion (p$_i \neq 0$). Upon evaluating the inversion probability (p$_i$), some devices are reversed, shown in green in \textbf{c)}.}
 \label{fig:sch_2} 
\end{figure*}

\section{Modeling of individual memristive units}

\noindent Efforts to describe a compact memristive model in SPICE are extensive and carried out by many groups~\cite{Biolek2009,Joglekar2009,Yakopcic2012,Patterson2017}. These models are usually based on the theoretical description by Chua~\cite{chua_memristor-missing_1971} and the empirical formulation by Strukov \textit{et al.}~\cite{strukov_missing_2008}. The associated mathematical equations~\cite{pershin_memory_2011} are: 1) generalized Ohm's law: $I (t) = R_M^{-1}(X, V_M, t) \cdot V_M(t)$ for the memristive resistance ($R_M$), and 2) the evolution of the internal state variable ($X$): $\frac{dX}{dt} = f(X, V_M, t)$ which describes the behaviour and memory of the memristive unit~\cite{pershin_memory_2011}. $R_M$ is a function of $X$, time ($t$), and the applied voltage ($V_M$).  

\noindent Among a considerable amount of models for memristive devices~\cite{pershin_validity_2020}, we have chosen the one proposed by Pershin and Ventra~\cite{pershin_spice_2013} which sets $R_M = X$ and defines $f(X, V_M, t) = f(V_M) = \beta \cdot [V_M - 0.5 \cdot (\lvert V_M + V_t \rvert - \lvert V_M - V_t \rvert)]$ modulated by a window function. Depending on the value of $\beta$, this model allows to mimic both threshold ($\lvert \beta \rvert >> 10^7$) and progressive ($\lvert \beta \rvert << 10^7$) switching \cite{PERSHIN202052}.  

NGSPICE is the engine chosen in this study to perform the aforementioned calculation. The memristive internal state ($X$) is computed by means of a subcircuit~\cite{pershin_spice_2013}, comprising a capacitor and a linear voltage-controlled current source, to which a modification by Vourkas and Sirakoulis~\cite{vourkas_memristor-based_2015} is introduced to effectively restrict the resistance values between two selected boundaries (R$_\mathrm{L}$ and R$_\mathrm{H}$).  

The individual model is thus bipolar\footnote{Negative polarity is expected to produce resistance reduction (SET-like operation) while the positive is programmed to increase the resistance (RESET-like).} and the parameters used for each memristive unit are the threshold voltage $V_t$, determining the onset of the resistance change; R$_\mathrm{L}$ and R$_\mathrm{H}$, low and high limits for $R_M = X$, respectively; and $\beta$ which sets a rate of change for $X$ depending on the simulation step. As a reference, $\beta < 10^7$ $\frac{\Omega}{V \cdot s}$ produces a moderate rate of change that upon a proper external stimulus' choice makes more likely to observe partial switching, reducing the switching ratio ($r$) compared to the maximum allowed $\left(r^{max} = \frac{\mathrm{R}_\mathrm{H}}{\mathrm{R}_\mathrm{L}}\right)$. 

\section{Collective electrical response of arrays}

\noindent Geometrical arrays of identical and reliable memristive units are built as indicated in Fig.~\ref{fig:sch_2}. A custom Python script \mbox{generates} the arrays as square-shaped networks of nodes connected by horizontal and vertical edges. Each edge contains a memristive unit whose polarity is uniformly initialized, as shown in Fig.~\ref{fig:sch_2}a. Such initialized homogeneous configuration can be distorted by either removing nodes (Fig.~\ref{fig:sch_2}b) and/or reversing individual unit's polarity (Fig.~\ref{fig:sch_2}c). Distortions are randomly introduced by two probabilities $p_r$ and $p_i$ for removal and reversal, respectively.

Once the network is set, the script generates a netlist to be simulated by NGSPICE in its batch mode. Each memristive unit is computed using the previously described model using exactly the same set of parameters (R$_\mathrm{L} = 2$~k$\Omega$, R$_\mathrm{H} = 200$~k$\Omega$, V$_t = 0.6$~V, and $\beta = 5\cdot10^5$~$\frac{\Omega}{V \cdot s}$), and - except stated otherwise - initialized in the same condition (R$_\mathrm{init}$ = R$_\mathrm{H}$).   

The electrical responses of the arrays are analyzed applying a voltage source between nodes (0,0) and (3,0) (see Fig.~\ref{fig:sch_2}a), and quantified by considering the current flow between them. \textbf{R}$_{coll}$ is a collective measure of each array's resistance state corresponding to the linear fit of the externally applied voltage as a function of the current, determined in the vicinity of $0$~V (remnant state). The applied stimuli comprise sinusoidal semi-cycles, each of which will be referred to as pulses.

\begin{figure}[ht!]
    \centering
    \includegraphics[width=8cm]{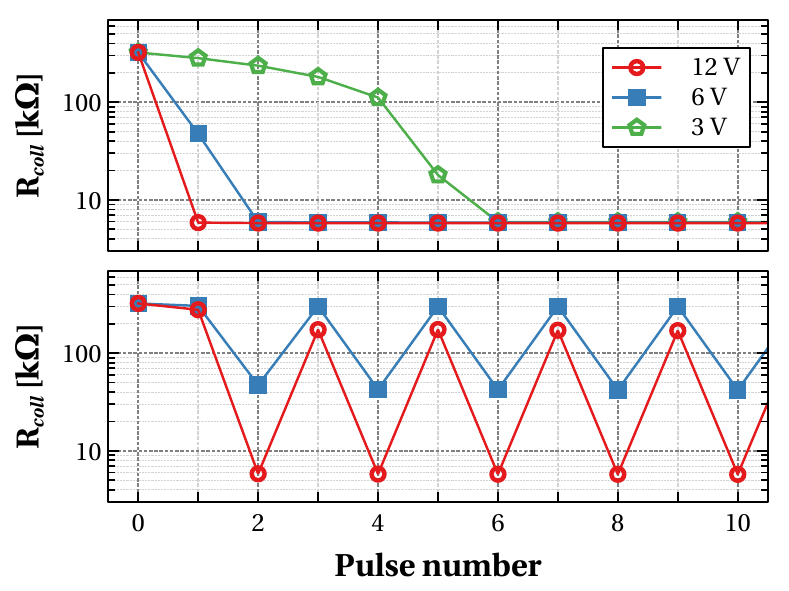}
    \caption{\textbf{R}$_{coll}$ of a homogeneous array, as the one shown in Fig.~\ref{fig:sch_2}a, as a function of the applied pulses. By initializing every memristive unit in R$_\mathrm{init} = 200$~k$\Omega$ the \textbf{upper panel} demonstrates the response produced upon multiple negative pulses of different amplitudes (A = 12, 6, and 3 V) repetitively applied while the \textbf{lower panel} displays the two possible states achieved when subjected to alternating polarity pulses of different amplitudes.}
 \label{fig:accumulation} 
\end{figure}

The first experiment explores arrays' accumulation ability. A homogeneous array of 4 x 4 nodes, as depicted in the upper panel of Fig.~\ref{fig:sch_2}, is defined. With each device initialized as R$_\mathrm{init} = 200$~k$\Omega$, a train of 10 negative-polarity pulses is applied. The collective resistance \textbf{R}$_{coll}$ is depicted in the upper panel of Fig.~\ref{fig:accumulation} as a function of the pulse number. Before applying any pulse, initializing each unit as R$_\mathrm{init} = 200$~k$\Omega$ corresponds to \textbf{R}$_{coll}$ $\sim$ $300$~k$\Omega$. Upon applying negative pulses \textbf{R}$_{coll}$ gets reduced until a limit is reached. How fast this limit is approached depends on the pulses' amplitude. This experiment demonstrates arrays' ability to display accumulation upon a proper choice of the external stimuli. 

To test if the achieved states are stable, a homogeneous array of the same characteristics as before is set and 10 pulses of alternating polarity are applied. The results are shown in the lower panel of Fig.~\ref{fig:accumulation}. When the array is cycled, \textbf{R}$_{coll}$ eventually reaches a steady condition switching back and forth between two values whose ratio ($r$) depends on the amplitude of the external signal. 

\vspace{-0.4cm}

\begin{figure}[ht!]
    \centering
    \includegraphics[width=8cm]{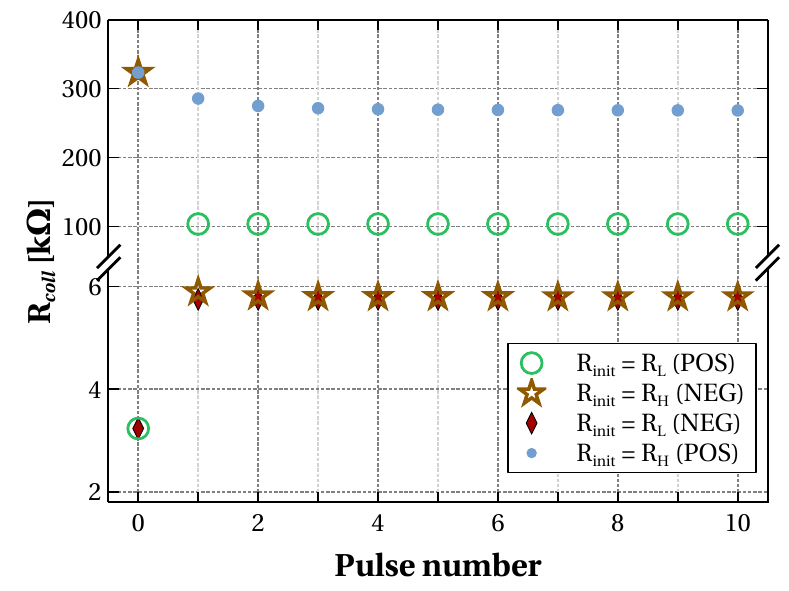}
    \caption{\textbf{R}$_{coll}$ of a homogeneous array, as the one shown in Fig.~\ref{fig:sch_2}a, as a function of the applied pulses of a fixed amplitude (A = 10 V). By initializing every memristive device homogeneously (same condition for all of them in the same run) despite the pulses' polarity, an irreversible change is always observed at first to then reach a steady condition.}
 \label{fig:forming} 
\end{figure}

\noindent A careful examination of the lower panel of Fig.~\ref{fig:accumulation} indicates an unexpected phenomenology that is further analyzed. Even when each device is homogeneously initialized in its highest resistance state (R$_\mathrm{H} = 200$~k$\Omega$), and the first applied pulse is positive (polarity for which individual units - already at R$_\mathrm{H}$ - are not expected to vary), a noticeable reduction of \textbf{R}$_{coll}$ is observed. To systematically address this issue, the same array as before is set and four simulations are run. Each of them corresponds to homogeneously initializing every memristive unit in R$_\mathrm{H}$ or R$_\mathrm{L}$ and applying either only positive or negative pulses. Fig.~\ref{fig:forming} demonstrates there exists, for every possible combination, a pristine state that can never be recovered. 

\vspace{-0.4cm}

\begin{figure}[ht!]
    \centering
    \includegraphics[width=8cm]{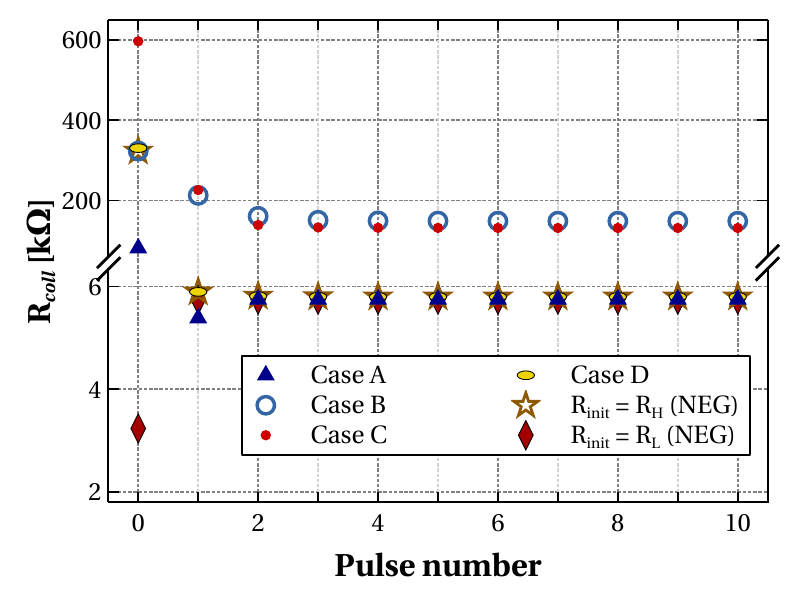}
    \caption{\textbf{R}$_{coll}$ of distorted arrays, as shown in Figs. \ref{fig:sch_2}b and c, as a function of the applied pulses (A = 10 V). \textbf{Case A} randomizes R$_\mathrm{init}$, \textbf{Case B} introduces $p_i = 0.5$ allowing to revert each memristive unit's polarity with a 50 $\%$ probability. \textbf{Cases C} and \textbf{D} remove (1,0) and (1,3) nodes, respectively.}
 \label{fig:distortions} 
\end{figure}

\noindent Until now, homogeneous and undistorted arrays were studied upon slightly different stimuli schemes. In the following, distortions are introduced to analyze the impact of each change in the collective response. Fig.~\ref{fig:distortions} includes the results of four experiments. \textbf{Case A} enables the initial state of each device to be randomly set as either R$_H$ or R$_L$. \textbf{Case B} allows the polarity of each memristive unit to be reversed with a 50 $\%$ probability (p$_i$ = 0.5). Finally, \textbf{Cases C} and \textbf{D} are the result of intentionally removing (p$_r$ $\neq$ 0) two particular nodes ((1,0) and (1,3)), one for each case. The distortions are introduced one at a time.  

\vspace{-0.5cm}

\vspace{-0.1cm}
\noindent Two cases display a different pristine state: \textbf{Case A}, when randomizing each memristive unit's initial state, and \textbf{Case C} when the node (1,0) is removed. In the first one, the pristine state demonstrates a lower resistance value (\textbf{Case A}) while in the other (\textbf{Case C}) is higher than the all-in-R$_H$ reference condition. The difference in the response obtained upon removal of nodes: (1,0) (\textbf{Case C}) and (1,3) (\textbf{Case D}) relies on the texture developed within the array due to the instantiated source. Since the external signal is sourced between (0,0) and (3,0), and at least when the array is homogeneous and undistorted, a preferential current path is formed along the first column~\cite{fdf2021} (easily understandable since it represents the shortest and less resistant path for the current to flow). If distortion is included in such a preferential path, the consequence is much more noticeable than if the removal had occurred outside of it (as is the case for (1,3)). Other distortions are noticeable in further cycling. For instance, when inversion is allowed, the polarity and intensity of voltage drop - required to fully switch each unit - get modified, and progressive changes are observed (even when the external amplitude is kept the same for every experiment, \textbf{Case B}). Moreover, the effective ratio ($r$) depends not only on the initial resistance state (consider \textbf{Case A}) but also on the polarity of each device (\textbf{Case B}) and the geometry of the array (\textbf{Cases C}, and \textbf{D}).         


\section{Discussion}

\noindent In this brief communication, 
we characterized assemblies of simulated memristive units - represented as geometrically-initialized arrays further distorted - as a whole by using a global quantity named after collective resistance. By starting with very simple and ordered (so-called homogeneous) assemblies of bipolar homogeneous memristive units, we demonstrated accumulation and `forming' at the array level. Accumulation is the ability to display multiple resistance levels, intermediate between the two programmed extremes, upon trains of multiple equivalent pulses. This is a feature of the individual units themselves when the applied signal describes a minor loop and the switch is incomplete and expresses itself as an arrays' property for the same reason. The ability to partially switch the arrays' resistance state - with an effective resistance ratio lower than the programmed one - is a consequence of subjecting some units to a minor loop. The voltage drop distribution within the arrays determines the occurrence of units' partial switching which translates into partial switching of the collective electrical responses. `Forming', the term used to refer to a first and irreversible change, is a characteristic observed for the arrays as a whole that is not present in the individual model (in the latter, every change is reversible). As such it can be seen as an emergent feature. This phenomenology is the consequence of a combination of each devices' polarity, the connectivity among the units within the array, and the fact that a homogeneously initialized array could never be a condition experimentally available except for in a very particular pristine situation.

Figs. \ref{fig:accumulation} and \ref{fig:forming} exemplify how setting highly idealized memristive arrays of moderate size already displays a degree of complexity that needs to be taken into account when characterizing self-assemblies to be technologically exploited. Moreover, the introduction of simple distortions has a considerable impact on the available resistance states and their evolution upon cycling. 

As a final remark, we want to point out that our simulation is far from realistic and/or representative of the current dimensions, of either memristive crossbars or the number of connections in experimental realizations of self-assemblies. On the contrary, we want to illustrate the complexity that arises from the enhanced connectivity among units and the impossibility of characterizing and controlling the dispersion of their individual responses (including memristive model, initial state, device-to-device, and even cycle-to-cycle variation) using a global macroscopic electrical characterization. Moreover, such characterization, if not properly conducted, may affect the arrays' resistance configuration in an irreversible way.

\bibliographystyle{./bibliography/IEEEtran}
\bibliography{./bibliography/IEEEabrv,./bibliography/IEEEexample}

\end{document}